# QUANTUM CONFORMATIONAL TRANSITION IN BIOLOGICAL MACROMOLECULE

LUO LiaoFu

*Faculty of Physical Science and Technology, Inner Mongolia University, Hohhot 010021, China*

email: lolfcm@mail.imu.edu.cn

**Abstract**

The conformational change of biological macromolecule is investigated from the point of quantum transition. A quantum theory on protein folding is proposed. Compared with other dynamical variables such as mobile electrons, chemical bonds and stretching-bending vibrations the molecular torsion has the lowest energy and can be looked as the slow variable of the system. Simultaneously, from the multi-minima property of torsion potential the local conformational states are well defined. Following the idea that the slow variables slave the fast ones and using the nonadiabaticity operator method we deduce the Hamiltonian describing conformational change. It is proved that the influence of fast variables on the macromolecule can fully be taken into account through a phase transformation of slow variable wave function. Starting from the conformation-transition Hamiltonian the nonradiative matrix element is calculated in two important cases: A, only electrons are fast variables and the electronic state does not change in the transition process; B, fast variables are not limited to electrons but the perturbation approximation can be used. Then, the general formulas for protein folding rate are deduced. The analytical form of the formula is utilized to study the temperature dependence of protein folding rate and the curious non-Arrhenius temperature relation is interpreted. The decoherence time of quantum torsion state is estimated and the quantum coherence degree of torsional angles in the protein folding is studied by using temperature dependence data. The proposed folding rate formula gives a unifying approach for the study of a large class problems of biological conformational change.

## 1 Molecular torsion as slow variable

There are huge numbers of variables in a biological system. What are the fundamental variables in the life processes at the molecular level? Since the classical works of B. Pullman and A. Pullman on nucleic acids [1], it is generally accepted that the mobile $\pi$ electrons play an important role in the biological activities of macromolecules. However, the traditional quantum biochemistry cannot treat a large class problems relating to the conformational variation of biological macromolecules such as protein folding, signal transduction and gene expression regulation, etc. In fact, for a macromolecule consisting of n atoms there are 3n coordinates if each atom is looked as a point. Apart from 6 translational and rotational degrees of freedom there are 3n-6 coordinates describing molecular shape. The molecular shape is the main variables responsible for conformational change. It has been proved that the bond lengths, bond angles and torsion (dihedral) angles form a complete set to describe the molecular shape.

For a complex system consisting of many dynamical variables the separation of slow/fast variables is the first key step in investigation. In his synergetics Haken proposed that the long-living systems slave the short-living ones, or briefly, the slow variables slave the fast ones. He indicated that the fast variables can be adiabatically eliminated in classical statistical mechanics [2]. However, what is the slow variable for a molecular biological system? The typical chemical bond energy is several electron volts (for example, 3.80 ev for C-H bond, 3.03 ev for C-N bond, 6.30 ev for C=O dissociation). The CG hydrogen bond energy is 0.2 ev and the TA hydrogen bond energy is 0.05 ev in nucleic acids. The energy related to the variation of bond length and bond angle is in the range of 0.4-0.03 ev. While the torsion vibration energy is 0.03-0.003 ev, the lowest in all forms of biological energies. In terms of frequency, the stretching and bending frequency is $10^{14}-10^{13}$ Hz while that for torsion is $7.5 \times 10^{12}-7.5 \times 10^{11}$ Hz. Interestingly, the torsion energy is even lower than the average thermal energy per atom at room temperature (0.04 ev in 25 C); the torsion angles are easily changed even at physiological temperature. Therefore, the torsion motion can be looked as the slow variable and others including mobile $\pi$ electron, chemical binding, stretching and bending etc are fast variables.

The torsion motion has two important peculiarities. The torsion energy 0.03-0.003 ev (0.1-1 Kcal/mole) corresponds to vibration frequency in the range of far-infrared spectrum. From the vibrational partition function of a molecule in harmonic conformational potential

$$Z = (e^{(1/2)\beta\hbar\omega} - e^{-(1/2)\beta\hbar\omega})^{-1} \tag{1}$$

one deduces the average energy $\overline{E}$ and entropy $S$ readily:

$$\overline{E} = \frac{\hbar\omega}{2} + \hbar\omega(e^{\beta\hbar\omega} - 1)^{-1} \tag{2}$$

$$S = k_B \{\beta\hbar\omega(e^{\beta\hbar\omega} - 1)^{-1} - \ln(e^{\beta\hbar\omega} - 1) + \beta\hbar\omega\} \tag{3}$$

For a molecule with many torsion angles the potential can be expanded into harmonic modes and each mode contributes to internal energy and entropy described by (2) and (3). As is well known, the Boltzmann entropy is related to Shannon information quantity by

$$S = (k_B \ln 2)I \tag{4}$$

Thus, by means of (3) one sees the strong dependence of information quantity on frequency. For example, as $T = 300$ K the information quantity is $I \sim 10^{-6}$ for frequency $v = 10^{14}$ Hz, but $I=0.63$ for $v = 10^{13}$, $I=2.83$ for $v = 10^{12}$, $I=4.14$ for $v = 10^{11}$ and $I=8.74$ for $v = 10^9$. As the frequency lower than $10^{13}$ Hz, the information quantity $I$ increases rapidly. This is the so-called Boson condensation. Since the typical value of torsion frequency is $10^{12}-10^{13}$ Hz, the torsion vibration may play an important role in the transmission of information in the biological macromolecular system.

The second peculiarity is, different from stretching and bending the torsion potential generally has several minima with respect to angle coordinate that correspond to several stable conformations. To study conformational dynamics from the point of quantum mechanics we should discuss the localization of conformational state first in the introduction.

To be definite, consider torsion potential with two minima. Set

$$U(\theta) = \begin{cases} \dfrac{\kappa}{2}(\theta + \dfrac{\pi}{2})^2 & [-\pi,\ 0] \\ \dfrac{\kappa}{2}\{(1-\lambda)\ (\theta - \dfrac{\pi}{2})^2 + \dfrac{\pi^2}{4}\lambda\} & [0,\ \pi] \end{cases} \quad (5)$$

Here $\lambda$ is an asymmetrical parameter. If $\lambda = 0$, then $U(\theta)$ is $C_2$ symmetrical, and the conformational wave function is nonlocal and no definite conformation can be related with the molecule. However if $\lambda \neq 0$, the $C_2$–symmetry is broken; by numerical solution of Schrodinger equation for the potential given by (5), one can show that for $I\kappa/\hbar^2 = 10$ ($I$ is the inertia moment of the molecule with respect to the coordinate $\theta$), the ground state would be localized to an amount larger than 90% as $\lambda \geq 10^{-3}$, and for $I\kappa/\hbar^2 = 40$, the ground state would be localized to larger than 99% as $\lambda \geq 10^{-6}$ [3]. Therefore, the small asymmetry in potential (which does exist for a real macromolecule) would cause the strong localization of wave functions. In his study on the excitations of a disordered lattice, Anderson proposed a noted theorem: the eigenfunctions are localized if the strength of the disorder exceeds some definite value [4]. Now, the theory of Anderson localization has been applied to the system of molecular conformations. It means that the localized quantum conformational state can well be defined for a biological macromolecule.

## 2 Conformational change as a quantum transition calculated by non-adiabatic operator method

We propose that the conformation variation of a macromolecule is a process of quantum transition between different torsion states. Apart from torsion coordinates, the transition is also related to some fast variables of the system, for example, the frontier electrons of the molecule, the stretching-bending of the molecule and the atomic group connected to the molecule, etc. So, the dynamical variables of the system are $(\theta, x)$ where $x$ describes the coordinates of fast variables and $\theta$ the torsion angles of the molecule. The wave function $M(\theta, x)$ satisfies

$$(H_1(\theta, \dfrac{\partial}{\partial \theta}) + H_2(x, \dfrac{\partial}{\partial x}; \theta))M(\theta, x) = EM(\theta, x) \quad (6)$$

$$H_1 = \sum -\dfrac{\hbar^2}{2I_j}\dfrac{\partial^2}{\partial \theta_j^2} + U(\theta) \quad (7)$$

Here $I_j$ denotes the inertial moment of the $j$-th torsion and the torsion potential $U$ is a function of a set of torsion angles $\theta = \{\theta_j\}$. Because the fast variables change more quickly than the variation of torsion angles, the adiabatic approximation can be used. In adiabatic approximation the wave function is expressed as

$$M(\theta,\ x) = \psi(\theta)\varphi(x,\ \theta) \quad (8)$$

and these two factors satisfy

$$H_2(x, \dfrac{\partial}{\partial x}; \theta)\varphi_\alpha(x,\ \theta) = \varepsilon_\alpha(\theta)\varphi_\alpha(x,\ \theta) \quad (9)$$

$$\{H_1(\theta, \frac{\partial}{\partial \theta}) + \varepsilon_\alpha(\theta)\}\psi_{kn\alpha}(\theta) = E_{kn\alpha}\psi_{kn\alpha}(\theta) \tag{10}$$

here $\alpha$ denotes the quantum number of fast-variable wave function $\varphi$, and $(k, n)$ refer to the conformational and the vibrational state of torsion wave function $\psi$, respectively.

Because $M(\theta, x)$ is not a rigorous eigenstate of Hamiltonian $H_1 + H_2$, there exists a transition between adiabatic states that results from the off–diagonal elements[5]

$$\int M^+_{k'n'\alpha'}(H_1 + H_2)M_{kn\alpha}d\theta dx = E_{kn\alpha}\delta_{kk'}\delta_{nn'}\delta_{\alpha\alpha'} + \langle k'n'\alpha' | H' | kn\alpha \rangle \tag{11}$$

$$\langle k'n'\alpha' | H' | kn\alpha \rangle = \int \psi^+_{k'n'\alpha'}(\theta)\sum_j -\frac{\hbar^2}{2I_j}\{\int \varphi^+_{\alpha'}(\frac{\partial^2 \varphi_\alpha}{\partial \theta_j^2} + 2\frac{\partial \varphi_\alpha}{\partial \theta_j}\frac{\partial}{\partial \theta_j})dx\}\psi_{kn\alpha}(\theta)d\theta \tag{12}$$

Here $H'$ is a Hamiltonian describing conformational transition. We see that the conformational transition is related to the fast-variable wave function $\varphi_a(x, \theta)$ and is determined by its $\theta$ dependence. Equation (12) is the generalization of the nonadiabaticity operator of Huang and Rhys in solid state physics (1950) [5]. The nonadiabatic matrix element (12) will be calculated in two important cases.

*Case* A  Protein folding with only electrons as fast variables and electronic state not changing ($\alpha' = \alpha$)

Consider protein folding with only electrons as fast variables. For most protein folding problem the electronic state does not change in transition processes, namely $\alpha' = \alpha$. In molecular orbital theory the electronic wave function $\varphi_a$ can be expressed as the linear combination of atomic orbits, and the combination coefficients and the eigenvalue of energy $\varepsilon_a(\theta)$ are obtained by solving Huckel equations. Because the wave function $\varphi_a$ is generally real, one can deduce

$$\int \varphi_a(x, \theta)\frac{\partial \varphi_a(x, \theta)}{\partial \theta}dx = 0 \tag{13}$$

from the normalization condition $\int \varphi_a(x, \theta)\varphi_a(x, \theta)dx = 1$. Therefore, only the first term in Eq. (12) is retained, namely

$$\langle k'n'\alpha | H' | kn\alpha \rangle = \int \psi^+_{k'n'\alpha}(\theta)\sum_j \{-\frac{\hbar^2}{2I_j}\int \varphi^+_\alpha \frac{\partial^2 \varphi_\alpha}{\partial \theta_j^2}d^3 x\}\psi_{kn\alpha}(\theta)d\theta \tag{14}$$

The case will be discussed in detail in section 4.

*Case* B  Protein folding in general but perturbation approximation applicable

In the case of fast variables not limited to the coordinates of electrons, $\alpha' \neq \alpha$ should be considered. The calculation is more complicate in this case but if the perturbation approximation can be used then the $\theta$ dependence of fast-variable wave function $\varphi_\alpha(x, \theta)$ can be deduced by the perturbation method as follows:

$$H_2(x, \frac{\partial}{\partial x};\theta) = H_2(x, \frac{\partial}{\partial x};\theta_0) + \sum_j (\frac{\partial H_2(x, \frac{\partial}{\partial x};\theta)}{\partial \theta_j})_0 (\theta_j - \theta_{j0})$$

$$\equiv H_2(x, \frac{\partial}{\partial x};\theta_0) + \sum_j h^{(j)}(x, \frac{\partial}{\partial x})(\theta_j - \theta_{j0})$$

$$h^{(j)}(x, \frac{\partial}{\partial x}) = (\frac{\partial H_2(x, \frac{\partial}{\partial x};\theta)}{\partial \theta_j})_0 \tag{15}$$

$$\varphi_\alpha(x, \theta) = \varphi_\alpha(x, \theta_0) + \sum_j (\theta_j - \theta_{j0}) \sum_{\beta \neq \alpha} \frac{h^{(j)}_{\beta\alpha}}{\varepsilon^{(0)}_\alpha - \varepsilon^{(0)}_\beta} \varphi_\beta(x, \theta_0) \tag{16}$$

Inserting (16) into (12), only the second term is retained and one has

$$\langle k'n'\alpha' | H' | kn\alpha \rangle$$
$$= \sum_j \frac{i\hbar}{\sqrt{I_j}} a^{(j)}_{\alpha'\alpha} \int \psi^+_{k'n'\alpha'}(\theta) \frac{\partial}{\partial \theta_j} \psi_{kn\alpha}(\theta) d\theta \tag{17}$$

$$a^{(j)}_{\alpha'\alpha} = \frac{i\hbar}{I_j^{1/2}} \int \varphi^+_{\alpha'}(x, \theta_0) \sum_\beta{}' \frac{h^{(j)}_{\beta\alpha}}{\varepsilon^{(0)}_\alpha - \varepsilon^{(0)}_\beta} \varphi_\beta(x, \theta_0) dx$$

$$= \frac{i\hbar}{I_j^{1/2}} \frac{h^{(j)}_{\alpha'\alpha}}{\varepsilon^{(0)}_\alpha - \varepsilon^{(0)}_{\alpha'}} (1 - \delta_{\alpha'\alpha}) \tag{18}$$

The amplitude $a^{(j)}_{\alpha'\alpha} \neq 0$ in first-order perturbation since $\alpha' \neq \alpha$ in this case. However, as $\alpha' = \alpha$ the amplitude $a^{(j)}_{\alpha'\alpha} = 0$ due to the orthogonality of the wave function and the effect of higher order perturbation should be considered. The further calculation of case B will be given in section 5.

## 3  Phase factor of torsion wave function due to coupling to fast variable

To obtain a rigorous solution of equation (6), namely, to solve the equation

$$(\sum_j -\frac{\hbar^2}{2I_j} \frac{\partial^2}{\partial \theta_j^2} + U(\theta) + H_2(x, \frac{\partial}{\partial x};\theta)) M(\theta, x) = EM(\theta, x) \tag{19}$$

without adiabatic approximation the total wave function $M(\theta, x)$ should be expanded as

$$M(\theta, x) = \sum_\alpha \psi_\alpha(\theta) \varphi_\alpha(x, \theta) \tag{20}$$

where $\varphi_\alpha(x, \theta)$ is the fast-variable wave function satisfying equation (9). Inserting (20) into (19) we obtain coupled equations for $\psi_\alpha(\theta)$

$$\sum_{\alpha'} \{(\sum_j -\frac{\hbar^2}{2I_j} \frac{\partial^2}{\partial \theta_j^2} + U(\theta) + \varepsilon_{\alpha'}(\theta))\delta_{\alpha'\alpha} + f_{\alpha\alpha'}(\theta)\} \psi_{\alpha'}(\theta) = E\psi_\alpha(\theta) \tag{21}$$

$$f_{\alpha\alpha'}(\theta) = \sum_j \{-\frac{\hbar^2}{2I_j}\int \varphi_\alpha^*(x,\theta)\frac{\partial^2}{\partial \theta_j^2}\varphi_{\alpha'}(x,\theta)dx \qquad (22)$$

$$-\frac{\hbar^2}{2I_j}\int 2\varphi_\alpha^*(x,\theta)\frac{\partial}{\partial \theta_j}\varphi_{\alpha'}(x,\theta)dx \frac{\partial}{\partial \theta_j}\}$$

Note that (22) is exactly the expression for nonadiabatic operator whose matrix element has been given by Eq (12). Neglecting the non-diagonal term in (21) in first-order approximation we obtain

$$\sum_j \{-\frac{\hbar^2}{2I_j}\frac{\partial^2}{\partial \theta_j^2} - \frac{\hbar^2}{2I_j}\int \varphi_\alpha^*(x,\theta)\frac{\partial^2}{\partial \theta_j^2}\varphi_\alpha(x,\theta)dx \qquad (23)$$

$$-\frac{\hbar^2}{2I_j}\int 2\varphi_\alpha^*(x,\theta)\frac{\partial}{\partial \theta_j}\varphi_\alpha(x,\theta)dx \frac{\partial}{\partial \theta_j}\}\psi_\alpha(\theta) = (E - U(\theta) - \varepsilon_\alpha(\theta))\psi_\alpha(\theta)$$

Eq (23) can be put in the form of Berry's phase [6]. Set Berry's connection

$$A_{\alpha j} = i\int \varphi_\alpha^*(x,\theta)\frac{\partial}{\partial \theta_j}\varphi_\alpha(x,\theta)dx \qquad (24)$$

Eq (23) can be rewritten as

$$\sum_j \{\frac{\hbar^2}{2I_j}(i\frac{\partial}{\partial \theta_j} + A_{\alpha j})^2\}\psi_\alpha(\theta) = (E - U(\theta) - \varepsilon_\alpha'(\theta))\psi_\alpha(\theta) \qquad (25)$$

$$\varepsilon_\alpha'(\theta) = \varepsilon_\alpha(\theta) + \sum_j \frac{\hbar^2}{2I_j}(A_{\alpha j}^2 - \int \frac{\partial \varphi_\alpha^*(x,\theta)}{\partial \theta_j}\frac{\partial \varphi_\alpha(x,\theta)}{\partial \theta_j}dx) \cong \varepsilon_\alpha(\theta) \qquad (26)$$

The second equality in Eq (26) is due to the additional term being a minor correction to $\varepsilon_\alpha(\theta)$, in the same order as the already neglected non-diagonal term $f_{\alpha\alpha'}(\alpha' \neq \alpha)$. Set

$$F_\alpha(\theta) = \exp\{i\sum_j \int^{\theta_j} A_{\alpha j}(\theta)d\theta_j\} \qquad (27)$$

One has

$$(\frac{\partial}{\partial \theta_j} - iA_{\alpha j}(\theta))F_\alpha(\theta)\psi_\alpha(\theta) = F_\alpha(\theta)\frac{\partial}{\partial \theta_j}\psi_\alpha(\theta)$$

$$(\frac{\partial}{\partial \theta_j} - iA_{\alpha j}(\theta))^2 F_\alpha(\theta)\psi_\alpha(\theta) = F_\alpha(\theta)\frac{\partial^2}{\partial \theta_j^2}\psi_\alpha(\theta)$$

or

$$F_\alpha(\theta)^{-1}(\frac{\partial}{\partial \theta_j} - iA_{\alpha j}(\theta))F_\alpha(\theta) = \frac{\partial}{\partial \theta_j}$$

$$F_\alpha(\theta)^{-1}(\frac{\partial}{\partial \theta_j} - iA_{\alpha j}(\theta))^2 F_\alpha(\theta) = \frac{\partial^2}{\partial \theta_j^2} \qquad (28)$$

With replacement

$$\psi_\alpha(\theta) = F_\alpha(\theta)\psi'_\alpha(\theta) = \exp\{i\sum_j \int^{\theta_j} A_{\alpha j}(\theta)d\theta_j\}\psi'_\alpha(\theta) \tag{29}$$

in Eq (25) and by multiplication of $F_\alpha(\theta)^{-1}$ on the left finally we have

$$\sum_j \frac{-\hbar^2}{2I_j}\frac{\partial^2}{\partial \theta_j^2}\psi'_\alpha(\theta) = (E - U(\theta) - \varepsilon_\alpha(\theta))\psi'_\alpha(\theta) \tag{30}$$

The fast variables have been removed in the equation of torsion wave function $\psi'_\alpha(\theta)$. As the fast-variable wave function is real, the normalization of $\varphi_\alpha(x,\theta)$ leads to $A_{\alpha j}$ vanishing for all $j$. It means the existence of fast variable of this kind does not change the torsion wave function at all. Therefore we prove that, to the first-order approximation, the influence of fast variables on the system has been fully taken into account through a unitary transformation, Eq 29. This can be looked as a formal demonstration of the principle "slow variables slaving fast variables" applicable in the protein folding problem. Moreover, the resulting Eq (30) takes the same form as Eq (10), the basic equation for torsion wave function in adiabatic approximation.

In case of two sets of fast variables, x and y, coupling to torsion angles $\{\theta_j\}$, instead of Eq (19) the basic equation is

$$(\sum -\frac{\hbar^2}{2I_j}\frac{\partial^2}{\partial \theta_j^2} + U(\theta) + H_{2f}(x, \frac{\partial}{\partial x}; \theta) + H_{2e}(y, \frac{\partial}{\partial y}; \theta))M(\theta, x) = EM(\theta, x) \tag{31}$$

Instead of Eq (20) we have

$$M(\theta, x, y) = \sum_{\alpha\sigma} \psi_{\alpha\sigma}(\theta)\varphi_\alpha(x, \theta)\chi_\sigma(y, \theta) \tag{32}$$

Here $\varphi_\alpha(x, \theta)$ satisfies Eq (9) (namely $H_2 = H_{2f}$ in Eq (9)) and $\chi_\sigma(y, \theta)$ satisfies a similar equation

$$H_{2e}\chi_\sigma(y, \theta) = \lambda_\sigma(\theta)\chi_\sigma(y, \theta) \tag{33}$$

By the same procedure we can prove

$$\sum_j \frac{-\hbar^2}{2I_j}\frac{\partial^2}{\partial \theta_j^2}\psi'_{\alpha\sigma}(\theta) = (E - U(\theta) - \varepsilon_\alpha(\theta) - \lambda_\sigma(\theta))\psi'_{\alpha\sigma}(\theta) \tag{34}$$

where $\psi'_{\alpha\sigma}(\theta)$ is $\psi_{\alpha\sigma}(\theta)$ multiplied by appropriate phase factor. The result shows that the torsion wave function satisfies a simple equation of vibrational type which is irrespective with fast variable coordinates even for several sets of fast variables.

## 4  Protein folding rate deduced from quantum conformational transition

We will calculate protein folding rate based on the quantum conformational transition theory. In this section the simple case (case A) of only electrons as fast variables is considered. We assume the electronic state remains unchanged in the folding. The dynamical variables of the system are $(\theta, x)$ where $x$ denotes the coordinates of electrons and $\theta = (\theta_1,...,\theta_N)$ — a group of dihedral angles responsible for the conformation change.[7] As stated in section 2, in adiabatic approximation the wave function of conformation-electron system $M(\theta, x)$ can be expressed as $M_{kn\alpha}(\theta, x) = \psi_{kn\alpha}(\theta)\varphi_\alpha(x,\theta)$. The conformational wave function of multi-torsion $\theta = \{\theta_1, ..., \theta_N\}$ is expressed as

$$\psi_{kn\alpha}(\theta) = \psi_{k_1,n_1,\alpha_1}(\theta_1)......\psi_{k_N,n_N,\alpha_N}(\theta_N) \tag{35}$$

where $\psi_{k_j,n_j,\alpha_j}(\theta_j)$ can be approximately expressed by a wave function of harmonic oscillator with quantum number $n_j$. Note that the harmonic potential has equilibrium position at $\theta_j = \theta_{kj}^{(0)}$ corresponding to the $k_j$-th minimum $E_{kj}$ ($k_j$=1,2,…) of the potential (see Fig 1).  Starting from Eq (14) the conformational transition rate can readily be deduced.

After thermal average over the initial states the transition rate takes the form

$$W = \frac{2\pi}{\hbar}\sum_{\{n\}}|\langle k'n'\alpha | H' | kn\alpha \rangle|^2 B(\{n\},T)\delta(\sum_j(n'_j\hbar\omega'_j - n_j\hbar\omega_j - \delta E_j)) \tag{36}$$

Here $\omega_j$ and $\omega'_j$ are potential parameters of the $j$-th mode in conformational state $k_j$ (initial state) and $k_j'$ (final state) respectively and $\delta E_j = E_{kj} - E_{k'j}$. $B(\{n\},T)$ denotes the Boltzmann factor in initial state

$$B(\{n\},T) = \prod_j B(n_j,T) = \prod_j e^{-n_j\beta\hbar\omega_j}(1 - e^{-\beta\hbar\omega_j}) \tag{37}$$

$$\beta = \frac{1}{k_B T}$$

After summing over final states we have

$$W = \frac{2\pi}{\hbar}\sum_{\{n\}}|\langle k'n'\alpha | H' | kn\alpha \rangle|^2 B(\{n\},T)\rho_E \tag{38}$$

Here $\rho_E$ means state density,

$$\rho_E = \frac{1}{\partial E_f / \partial N_f} = \frac{1}{\hbar\bar{\omega}'}, \qquad N_f = \sum_j n'_j$$
$$E_f = \sum_j(n'_j\hbar\omega'_j + E_{k'j}) \tag{39}$$

$\bar{\omega}'$ is the average of $\omega'_j$ over $j$.

If only one torsion angle participates in the conformational transition we call it single-mode or single torsion transition; if several torsion angles participate simultaneously in one step of conformational transition then we call it multi-mode or multi-torsion transition. Consider single-mode case at first. In this case the subscript $j$ will be dropped. The transitional rate

$$W = \frac{2\pi}{\hbar^2 \omega'} I_E I_V \tag{40}$$

$$I_E = \left| \frac{-\hbar^2}{2I} \int \varphi_\alpha \frac{\partial^2}{\partial \theta^2} \varphi_\alpha d^3 x \right|^2_{\theta=\theta_0} \tag{41}$$

$$I_V = \sum_n \left| \int \psi^+_{k'n'\alpha}(\theta) \psi_{kn\alpha}(\theta) d\theta \right|^2 B(n,T) \tag{42}$$

$\theta_0$ means the torsion-angle coordinate taking a value of the largest overlap region of vibrational functions. Eq. (41) can be estimated roughly by the square of rotational kinetic energy of the electron during conformational change. The quantum number $n'$ in (42) is determined by $n$ through energy conservation. The overlap integral in Eq (42) was first calculated for the case of same frequency for initial and final states ($\omega=\omega'$). In this case after thermal average it gives [8]

$$I_V = \left(\frac{\bar{n}+1}{\bar{n}}\right)^{p/2} J_p(2Q\sqrt{\bar{n}(\bar{n}+1)}) e^{-Q(2\bar{n}+1)} \tag{43}$$

in which

$$\bar{n} = (e^{\beta\hbar\omega} - 1)^{-1}$$

$$Q = I\omega(\delta\theta)^2 / 2\hbar, \qquad p = \frac{\delta E}{\hbar\omega} \tag{44}$$

$\delta\theta = \theta_k^{(0)} - \theta_{k'}^{(0)}$ is the angular displacement and $\delta E = E_k - E_{k'}$ the energy gap between initial and final states. $J_P$ denotes the modified Bessel function and here $p$ is related to the net change in oscillator quantum number. The modified Bessel function is introduced from the expansion

$$\exp\left\{\frac{x}{2}\left(y + \frac{1}{y}\right)\right\} = \sum_{n=-1}^{\infty} y^n J_n(x) \tag{45}$$

By use of the asymptotic formula for Bessel function [9]

$$e^{-z} J_p(z) = (2\pi z)^{-1/2} \exp(-p^2/2z) \qquad \text{for} \quad z \gg 1 \tag{46}$$

$I_V$ can be further simplified. Taking $\bar{n} > 1$ (equivalent to $\hbar\omega < 0.69 k_B T$ which is satisfied for the typical torsion frequency) into account, Eq (43) can be simplified to

$$I_V = (2\pi z)^{-1/2} \exp\left(-\frac{p^2}{2z}\right) \exp\frac{\delta E}{2k_B T} \tag{47}$$

with

$$z = (\delta\theta)^2 \frac{k_B T}{\hbar^2} I \tag{48}$$

Note that for typical value $(\delta\theta)^2 = 0.01$, $I = 10^{-37}$ g.cm$^2$ one has $z = 40$ and the condition $z \gg 1$ is fulfilled.

For multi-torsion transition ($N$ modes), from Eq (37) to Eq (39) by using the same deduction one obtains

$$W = \frac{2\pi}{\hbar^2 \overline{\omega}} I_E I_V$$

$$= \frac{2\pi}{\hbar^2 \overline{\omega}} I_E \sum_{\{p_j\}} \prod_j I_{Vj} \qquad (49)$$

$$I_E = \left| \sum_j^M \frac{-\hbar^2}{2I_j} \int \varphi_\alpha \frac{\partial^2}{\partial \theta_j^2} \varphi_\alpha d^3 x \right|^2_{\theta_j = \theta_{j0}} \qquad (50)$$

$$I_{Vj} = \exp\{-Q_j(2\overline{n}_j + 1)\} \left(\frac{\overline{n}_j + 1}{\overline{n}_j}\right)^{p_j/2} J_{p_j}(2Q_j \sqrt{\overline{n}_j(\overline{n}_j + 1)}) \qquad (51)$$

with

$$\overline{n}_j = (e^{\beta\hbar\omega_j} - 1)^{-1}$$

$$Q_j = I_j \omega (\delta\theta_j)^2 / 2\hbar$$

$$p_j = \frac{\delta E_j}{\hbar \omega_j} \qquad (52)$$

and $p_j$ representing the net change in phonon number for oscillator mode $j$ and satisfying the constraint

$$\sum_j^N p_j = p = \Delta E / \hbar\overline{\omega} \qquad \Delta E = \sum_j^N \delta E_j \qquad (53)$$

($\overline{\omega}$ - the average of $\omega_j$, $\sum_j \hbar\omega_j p_j = \hbar\overline{\omega} p$) in the summation of Eq (49). Note that Eq (51) takes the same form as (43).

To simplify $I_V = \sum_{\{p_j\}} \prod_j I_{Vj}$ we use

$$\sum_{p_1 + \ldots + p_n = p} \exp(-z_1) J_{p_1}(z_1) \exp(-z_2) J_{p_2}(z_2) \ldots \exp(-z_n) J_{p_n}(z_n)$$

$$= \frac{1}{\sqrt{2\pi}} \frac{1}{\sqrt{z_1 + \ldots + z_n}} \exp\left\{-\frac{p^2}{2(z_1 + \ldots + z_n)}\right\} \qquad (54)$$

that was proved in [10]. Then we obtain

$$I_V \equiv \sum_{\{p_j\}} \prod_j I_{Vj} = \frac{1}{\sqrt{2\pi}} \exp(\frac{\Delta E}{2k_B T})(\sum_j^N z_j)^{-\frac{1}{2}} \exp(-\frac{p^2}{2\sum_j^N z_j}) \qquad (55)$$

Here

$$z_j = (\delta\theta_j)^2 \frac{k_B T}{\hbar^2} I_j \qquad (56)$$

with the same form as Eq (48) in single-mode case.

By using

$$I_E = \left\{ \sum_j^M \frac{\hbar^2}{2I_j} \langle l_j^2 \rangle \right\}^2$$

$$= \frac{\hbar^4}{4} (\sum_j^M \frac{a_j}{I_j})^2 \qquad a_j = \langle l_j^2 \rangle \qquad (57)$$

where $l_j$ = the $j$-th magnetic quantum number (with respect to $\theta_j$) of electronic wave function $\varphi_\alpha(x,\{\theta\})$, $a_j = <l_j^2> \approx O(1)$, a number in the order of magnitude of 1, and inserting (55) into (49) we obtain the final results of multi-mode transition in equal frequency case. Note that in Eqs (50) and (57) the summation is taken over $M$ in $N$ torsion modes since only a part of torsion angles are correlated with electronic wave function ($M \leq N$). The final result is

$$W = \frac{\hbar^3 \sqrt{\pi}}{2\sqrt{2}\delta\theta\bar{\omega}'} \exp\{\frac{\Delta E}{2k_B T}\} \exp\{\frac{-(\Delta E)^2}{2\bar{\omega}^2(\delta\theta^2 k_B T \sum_j^N I_j)}\} (k_B T)^{-1/2} (\sum_j^N I_j)^{-1/2} (\sum_j^M \frac{a_j}{I_j})^2$$

(58)

Here $\delta\theta = \sqrt{<(\delta\theta_j)^2>_{av}}$ is the average of angular shift over $N$ torsion modes.

In above calculation of overlap integral the same frequency for initial and final states ($\omega_j = \omega'_j$) has been assumed, now we will generalized the results to the case of non-equal frequencies between initial and final states. $\Delta E$ in Eq (58) means the free energy decrease of torsion vibration in the folding. From the statistical physics the free energy of a system of oscillators is expressed by[11]

$$G = \frac{1}{\beta} \sum_j \{\ln(1-\exp(-\beta\omega_j)) + E_j\} \qquad (59)$$

where $E_j$ means the potential minimum of the $j$-th oscillator. As the frequency shifts from $\omega_j$ to $\omega_j'$ the free energy variation

$$\delta_\omega G = \sum_j \int_{\omega_j}^{\omega'_j} \frac{\partial G}{\partial \omega_j} d\omega_j = \sum_j \int_{\omega_j}^{\omega'_j} \frac{\hbar}{\exp(\beta\hbar\omega_j)-1} d\omega_j$$

$$\cong \sum_j \frac{1}{\beta} \ln \frac{\omega_j'}{\omega_j} \quad (\beta\hbar\omega_j \ll 1) \tag{60}$$

Therefore, the free energy difference between torsion initial state (frequency $\{\omega_j\}$) and final state (frequency shifted to $\{\omega_j'\}$) is

$$\Delta G = -\delta_\omega G + \sum_j \delta E_j = \Delta E + \sum_j \frac{1}{\beta} \ln \frac{\omega_j}{\omega_j'}$$
$$= \Delta E + \lambda k_B T \tag{61}$$

($\lambda = \sum_{j=1}^{N} \ln \frac{\omega_j}{\omega_j'}$). Considering that the contribution of frequency variation to the folding rate comes mainly from the torsion vibration term, $\Delta E$ in the rate equation Eq (58) should be replaced by $\Delta G$ (Eq 61) for non-equal frequency case. Finally we obtain the folding rate [10]

$$W = \frac{\hbar^3 \sqrt{\pi}}{2\sqrt{2}\delta\theta\bar{\omega}'} \exp\{\frac{\Delta G}{2k_B T}\} \exp\{\frac{-(\Delta G)^2}{2\bar{\omega}^2(\delta\theta^2 k_B T \sum_j^N I_j}\} (k_B T)^{-1/2} (\sum_j^N I_j)^{-1/2} (\sum_j^M \frac{a_j}{I_j})^2 \tag{62}$$

From Eq (62) the unfolding rate $W(\text{unfolding})$ for the reversed process is easily obtained by the replacement of $\Delta G$ by $-\Delta G$ and $\bar{\omega}'(\bar{\omega})$ by $\bar{\omega}(\bar{\omega}')$ in $W(\text{folding})$. Thus we have

$$\ln\{\frac{W(\text{folding})}{W(\text{unfolding})}\} = \frac{\Delta G}{k_B T} + \frac{(\Delta G)^2}{2k_B T \varepsilon}(\frac{\bar{\omega}^2 - \bar{\omega}'^2}{\bar{\omega}'^2}) + \ln \frac{\bar{\omega}}{\bar{\omega}'} \tag{63}$$

($\varepsilon = \bar{\omega}^2 (\delta\theta)^2 \sum_j^N I_j$). Eq (63) means the condition of dynamical balance $W(\text{folding}) = W(\text{unfolding})$ is slightly different from the usual equilibrium condition for chemical reaction $\Delta G = 0$ due to the different bias samplings in frequency space of $\{\omega_j\}$ and $\{\omega'_j\}$, $\bar{\omega} \neq \bar{\omega}'$.

Notice that the relation between $W(\text{folding})$ and $W(\text{unfolding})$ can not result from the detailed balance theorem of quantum mechanics since the thermal average over initial states has not been considered in the theorem.

## 5  Protein folding rate in general

So far we have discussed the globular protein folding with only electronic coordinates as the fast variable. In the general case of protein structural variation many other fast variables should be considered. For example, in the conformational transition of tubulin heterodimer in neuronal microtubule, in the ligand-receptor binding for the membrane protein in the cellular signal transduction, and in the atomic-group binding to the residues in histone modification in the gene expression regulation, the fast variables may include the atomic-group or ligand binding to polypeptide chain, the hydrophobic interaction between subunits, the isomerization under external action, etc. Moreover, the stretching and bending of single bond may serve as fast variable accompanying torsion transition. All these fast-variables $\{x\}$ make contribution to the protein structural variation through a transition from the wave function $\varphi_\alpha(x)$ to $\varphi_{\alpha'}(x)$ ($\alpha' \neq \alpha$). To solve the problem we assume the perturbation approximation can be used. Starting from Eq (17) and (18) and inserting them into Eq (38) the transition rate is deduced, $W = W_{dia} + W_{ndi}$ ($W_{dia}$ - the diagonal part and $W_{ndi}$ - the non-diagonal part) [12]

$$W_{dia} = 2\pi\hbar \{\sum_{\{n\}}\sum_j^M \frac{1}{I_j}|a_{\alpha'\alpha}^{(j)}|^2 \left|\int \psi_{k'n'\alpha'}^+(\theta)\frac{\partial}{\partial\theta_j}\psi_{kn\alpha}(\theta)d\theta\right|^2 B(\{n\},T)\rho_E \quad (64)$$

$$W_{ndi} = 2\pi\hbar \{\sum_{\{n\}}\sum_{j\neq l}^M \frac{(a_{\alpha'\alpha}^{(j)})^* a_{\alpha'\alpha}^{(l)}}{\sqrt{I_j I_l}} \times$$

$$(\int \psi_{k'n'\alpha'}^+(\theta)\frac{\partial}{\partial\theta_j}\psi_{kn\alpha}(\theta)d\theta)^* \int \psi_{k'n'\alpha'}^+(\theta)\frac{\partial}{\partial\theta_l}\psi_{kn\alpha}(\theta)d\theta\} B(\{n\},T)\rho_E$$

(65)

where $\theta = \{\theta_1,...\theta_N\}$, $\{n\}=\{n_1,...,n_N\}$ and $M$ means the number of torsion angles correlated to fast variables. The calculation of overlap integral of vibrational wave function with derivative $\frac{\partial}{\partial\theta_j}$ can be performed by using the phonon annihilation/production operator

$$\xi_j = (\frac{I_j\omega_j}{2\hbar})^{1/2}(\theta_j + \frac{\hbar}{I_j\omega_j}\frac{\partial}{\partial\theta_j}), \quad \xi_j^+ = (\frac{I_j\omega_j}{2\hbar})^{1/2}(\theta_j - \frac{\hbar}{I_j\omega_j}\frac{\partial}{\partial\theta_j}) \quad (66)$$

which satisfies $[\xi_j, \xi_k^+] = \delta_{jk}$, and

$$-\frac{\hbar^2}{2I_j}\frac{\partial^2}{\partial\theta_j^2} + \frac{I_j\omega_j^2}{2}\theta_j^2 = (\xi_j^+\xi_j + \frac{1}{2})\hbar\omega_j = (n_j + \frac{1}{2})\hbar\omega_j$$

For single mode case when $\omega = \omega'$ we deduce the transition rate

$$W = \frac{\pi}{\hbar}|a_{\alpha'\alpha}|^2 \{(\overline{n}+1)I_V(p-1) + \overline{n}I_V(p+1)\} \quad (67)$$

$I_V$ is given by (43) or its simplified form (47). For multi-mode case we deduce

$$W_{dia} = \frac{\pi}{\hbar}\sum_j^M |a_{\alpha'\alpha}^{(j)}|^2 \{(\overline{n}_j+1)I_{Vj}(p_j-1)\sum_{\{p_l\}}\prod_{l\neq j}^N I_{Vl}(p_l)$$

$$+\overline{n}_j I_{Vj}(p_j+1)\sum_{\{p_l\}}\prod_{l\neq j}^N I_{Vl}(p_l)\}$$

(68)

With aid of Eq (54) the above equation can be rewritten in a simplified form

$$W_{\text{dia}} = \frac{\pi}{\hbar} \sum_{j}^{M} \left| a_{\alpha'\alpha}^{(j)} \right|^2 (\{(\bar{n}_j + 1) I_V(p-1) + \bar{n}_j I_V(p+1)\} \tag{69}$$

where $I_V(p)$ is given by Eq (55) (denoted as $I_V$ there) and

$$I_V(p \pm 1) = \frac{1}{\sqrt{2\pi}} \exp(\frac{\Delta E}{2k_B T})(\sum_{j}^{N} Z_j)^{-\frac{1}{2}} \exp(-\frac{(p \pm 1)^2}{2\sum_{j}^{N} Z_j}) \tag{70}$$

Sarai and Kakitani studied radiationless transitions of molecules with large nuclear rearrangement [13]. By using generating function method [14] they deduced a similar formula but no clear analytical expression was obtained. Notice that in their formula of $W_{\text{dia}}$ the additional terms

$$Q_j \{(6\bar{n}_j^2 + 6\bar{n}_j + 1) H(0) - 2(\bar{n}_j + 1)(2\bar{n}_j + 1) H(\hbar\omega_j) -$$
$$-2\bar{n}_j(2\bar{n}_j + 1) H(-\hbar\omega_j) + (\bar{n}_j + 1)^2 H(2\hbar\omega_j) + \bar{n}_j^2 H(-2\hbar\omega_j)\})$$

appeared but we can prove that these terms equal zero due to $H(\pm\hbar\omega_j) \cong H(\pm 2\hbar\omega_j) \cong H(0)$ and their result is essentially identical with ours.

The non-diagonal term $W_{\text{ndi}}$ calculated by generating function method gives [13]

$$W_{ndi} = \sum_{j \neq l} \frac{\pi}{\hbar} (a_{\alpha'\alpha}^{(j)})^* a_{\alpha'\alpha}^{(l)} \sqrt{Q_j Q_l}$$
$$\{(2\bar{n}_j + 1)(2\bar{n}_l + 1) H(0) + 2(\bar{n}_j + 1)\bar{n}_l H(\hbar\omega_j - \hbar\omega_l) \tag{71}$$
$$-2(2\bar{n}_j + 1)(\bar{n}_l + 1) H(\hbar\omega_l) - 2\bar{n}_j(2\bar{n}_l + 1) H(-\hbar\omega_j)$$
$$+(\bar{n}_j + 1)(\bar{n}_l + 1) H(\hbar\omega_j + \hbar\omega_l) + \bar{n}_j \bar{n}_l H(-\hbar\omega_j - \hbar\omega_l)\}$$

Here $H(x)$ is the Fourier transformation of phonon-generating function, expressed as

$$H(x) = \int dt \exp\{-i\hbar^{-1}(\Delta E - x)t)\} G(t)$$

Due to the number $N$ of torsion modes cooperatively participating in the transition not small, $\Delta E \gg \hbar\omega_j$, all factors $H(x)$ in Eq (71) are essentially the same. Thus we have

$$W_{ndi} = \sum_{j \neq l} \frac{\pi}{\hbar} (a_{\alpha'\alpha}^{(j)})^* a_{\alpha'\alpha}^{(l)} \sqrt{Q_j Q_l} (-3\bar{n}_j + 3\bar{n}_l) H(0) \tag{72}$$
$$\cong 0$$

as the difference between $\bar{n}_j$ and $\bar{n}_l$ can be neglected $(\frac{\bar{n}_j - \bar{n}_l}{\bar{n}_j} = \frac{\omega_l - \omega_j}{\omega_l})$. So the total rate $W$ is given by the diagonal part, Eq (69).

Since the net variation $p$ of phonon number in torsion transition is much larger than 1 on account of the torsion mode number $N$ large enough we have $I_V(p) \cong I_V(p \pm 1)$ and Eq (69) is further simplified to

$$W = W_{\text{dia}} = \frac{\pi}{\hbar} \sum_j^M \left|a_{\alpha'\alpha}^{(j)}\right|^2 \{(2\bar{n}_j + 1)I_V(p)\}$$

$$\cong \frac{2\pi}{\hbar^2 \bar{\omega}'} M\bar{a}^2 k_B T I_V \tag{73}$$

where $\bar{a}^2$ means the average of $\left|a_{\alpha'\alpha}^{(j)}\right|^2$ over $j$ and $\overline{n_j} \approx \frac{k_B T}{\hbar \omega_j} \gg 1$ is used in the last equality.

In above deduction the same frequency for initial and final states has been assumed. However, the result can be generalized to the case of non-equal frequencies $\omega_j \neq \omega_j'$ through the replacement of $\Delta E$ by $\Delta G$ (Eq (61)) in the final equation (73). By using the asymptotic formula for Bessel function (Eq 46), Eq (73) can be written in a form similar to Eq (62) in section 4, namely

$$W = \frac{2\pi}{\hbar^2 \bar{\omega}'} I_V' I_E'$$

$$I_V' = \frac{\hbar}{\sqrt{2\pi}\delta\theta} \exp\{\frac{\Delta G}{2k_B T}\} \exp\{\frac{-(\Delta G)^2}{2\bar{\omega}^2(\delta\theta^2 k_B T \sum_j^N I_j)}\}(k_B T)^{1/2}(\sum_j^N I_j)^{-1/2}$$

$$I_E' = \sum_j^M \left|a_{\alpha'\alpha}^{(j)}\right|^2 \cong M\bar{a}^2$$

(74)

Here $I_V'$ is obtained by $I_V$ (Eq (55) with replacement of $\Delta E$ by $\Delta G$) multiplied by $k_B T$. Eq (74) can be used for unfolding as well as for folding and the relation between two rates is given by Eq (63).

## 6  Application and discussion : temperature dependence of protein folding rate

The temperature dependence of the transition rate reflects the essence of the folding dynamics of a protein. The comparison between theoretical calculation and experimental data on the temperature dependence of the rate will be able to give deeper insight into the folding mechanism. The theoretical formula for the folding rate has been given in Eq (62) or (74). To obtain quantitative result one should calculate the number of torsion modes $N$ first. N describes the coherence degree of multi-torsion transition in the protein folding. Based on the idea that the two-state protein folding is equivalent to a quantum transition between conformational states we assume that $N$ is obtained by the numeration of all main-chain and side-chain dihedral angles on the polypeptide chain except those residues on the tail of the chain which does not belong to any contact. A contact is defined by a pair of residues at least four residues apart in their primary sequence and with their spatial distance no greater than some threshold (say, 0.65 nm in the following study). Suppose the length of the chain (after removing residues on the tail) being $n$ and the number of the $i$-type residue on it being $n_i'$, the total number of torsion modes $N$ for a given

polypeptide chain is

$$N = 2n + \sum_i n'_i \qquad (i = 1, 2, \cdots, 20) \tag{75}$$

The side-chain dihedral angle number $n'_i$ takes 0 to 4 for different residues [15, 16].

As seen from Eq (62) and (74), to find the temperature dependence of folding rate one should know the relation between $\Delta G$ and temperature first. By using the relation of $\Delta G$ with $\Delta E$ (Eq (61)) and by consideration of the torsion potential susceptible to temperature near melting temperature $T_c$ (where a protein may undergo a transition of structure)

$$\Delta E(T) = \Delta E(T_c) + m(T - T_c) \tag{76}$$

we assume a linear relation between free energy change $\Delta G$ and temperature. This linearity has been tested rigorously by experiments [17]. Set

$$\eta = \frac{\Delta E(T_c) - mT_c}{\Delta E(T_c)}. \tag{77}$$

Assuming that 1) the measured value of folding free energy decrease is denoted by $\Delta G_f$ and 2) the measurement is carried out at temperature $T_f$, then one has

$$\Delta G_f = \Delta E(T_c)\{\eta + (1-\eta)T_f / T_c\} + k_B T_f \lambda. \tag{78}$$

where

$$\lambda = \sum_j \ln \frac{\omega_j}{\omega'_j} \cong N \ln \frac{\bar{\omega}}{\bar{\omega}'} \tag{79}$$

Inserting above equations into (62) or (74) we obtain the temperature dependence of logarithm rate [17][18]

$$\ln W(T) = \frac{S}{T} - RT \mp \frac{1}{2}\ln T + const. \qquad \text{(symbol - for \textit{case} A and + for \textit{case} B)} \tag{80}$$

where $const$ means temperature-independent term and

$$S = \frac{\eta \Delta E(T_c)}{2k_B}(1 - \frac{\eta \Delta E(T_c)}{\varepsilon}) \tag{81}$$

$$R = \frac{k_B}{2\varepsilon}(\lambda + \frac{m}{k_B})^2 = \frac{k_B}{2\varepsilon}(\lambda + \frac{(1-\eta)\Delta E(T_c)}{k_B T_c})^2$$

$$= \frac{1}{2\varepsilon k_B T_f^2}(\Delta G_f - \eta \Delta E(T_c))^2 \tag{82}$$

with

$$\varepsilon = \bar{\omega}^2 (\delta \theta)^2 \sum I_j = NI_0 \bar{\omega}^2 (\delta \theta)^2 \tag{83}$$

Finally we have

$$\frac{d \ln W}{d(\frac{1}{T})} = S + \frac{1}{2}T + RT^2 \qquad (Case \text{ A}) \tag{84A}$$

$$\frac{d\ln W}{d(\frac{1}{T})} = S - \frac{1}{2}T + RT^2 \qquad (Case\ B) \qquad (84B)$$

Note that Eq (80) for case A differs from case B only by an additive term $\frac{1}{2}\ln T$. This makes the difference of $\ln W(T+\Delta T) - \ln W(T)$ in two cases ($\Delta T \approx 50°$ for the full range of the measured temperature in experiments) being $\ln(1+\frac{\Delta T}{T})$ only, which is about 1.5% of the observed $\ln k_f$. Therefore the temperature dependence of folding rate deduced from two models are basically same. The reason is the temperature dependences of folding rate caused mainly by torsion motion, but not by fast-variables.

The temperature dependences of the folding and the unfolding rates can be studied on the same foot in this theory. Consider a folding whose slope parameters are denoted by $S$ and $R$, and its corresponding unfolding (under the same denaturant concentration and temperature) with slope parameters $S'$ and $R'$. From Eqs (81) and (82) we obtain a relation between them

$$\frac{R'}{R} = \frac{2k_B S' + \eta \Delta E(T_c)}{2k_B S - \eta \Delta E(T_c)} = \frac{\overline{\omega}^2}{\overline{\omega}'^2} \qquad (85)$$

The experiments on rate – temperature relationships in protein folding exhibit the following characteristics of non-Arrhenius behavior[19]. The folding rate decreases upon increase in temperature and even the crossover occurs at high temperature from normal positive barrier to abnormal negative. These characteristics can be explained by temperature–dependent terms in Eq (80). The last term $RT^2$ in (84A) (84B) is the main term contributed to the curvature of Arrhenius plot. Another law is: the plots of $\ln W$ versus $1/T$ are strongly curved for refolding of some proteins but almost linear for their unfolding under denaturant. This can be explained by the folding initial (unfolding final) frequency $\overline{\omega}$ lower than the unfolding initial (folding final) frequency $\overline{\omega}'$ for the studied proteins.

To make more quantitative comparison between theory and experiments Lu and Luo [17] [18] studied 16 proteins for which the experimental data on temperature dependence of rate and on folding free energy are currently available. They found that Eqs (84A) and (84B) are in good agreement with the rate-temperature relation for each protein. Moreover, through solving Eqs (81) and (82), one obtains $\eta\Delta E(T_c)$ and $\varepsilon$ (or $\delta\theta\overline{\omega}$) for each protein since the two slope parameters, $S$ and $R$, have been determined by temperature-dependent folding rates, and the free energies $\Delta G_f$ have been measured at some temperature $T_f$. Therefore, by using $S$, $R$ and the equilibrium free energy as input, the torsion potential parameters for each protein can be calculated. They found that the solutions of torsion potential parameters for 16 proteins are consistent with each other (Table 1). Moreover, in this approach the mutants exhibiting very different temperature dependencies than wild types were also interpreted. For example, the Arrhenius plot of wild-type λ-repressor (1lmb) shows ln$W$ decreasing with $1/T$ while that of λ-mutant G46A increasing with $1/T$ in the same temperature interval of 1000/$T$ from 3.0 to 3.2 [20].

In our studies the number of torsion modes $N$ is calculated from Eq (75) and it takes a value larger than 100 for most polypeptide chains. This means the coherence degree is generally large, consistent with the idea that many torsion angles participate in a quantum transition cooperatively in protein folding. We study how the experimental temperature dependence of folding rate gives a constraint in the determination of coherence degree *N*. Since the torsion parameters $\eta \Delta E(T_c)$ and $\varepsilon$ have been calculated from the experimental slope parameters *S* and *R*, one easily finds $\bar{\omega}^2$ is inversely proportional to *N* from the expression of $\varepsilon$, Eq (83). The maximal torsion frequency observed in experiments is about $7.5 \times 10^{12}$ Hz. This gives the minimal allowable *N*,

$$N_{th} = \text{Itg} \frac{\varepsilon}{I_0 (\delta\theta)^2 \bar{\omega}_{\max}^2}$$

where Itg x denotes the minimal integer larger than x. The slope parameters *S* and *R*, torsion parameter $\bar{\omega}\delta\theta$, coherence degree *N* and its threshold $N_{th}$ for 16 proteins are listed in Table 1. It is found that nearly all of these proteins have $N_{th}>1$. Therefore, to explain the temperature dependence of folding rate the introduction of coherent multi-torsion transition is necessary.

    In recent years the role of quantum decoherence was widely recognized by physicists [21-22]. Due to quantum entanglement with the environment the coherence or the ordering of the phase angles is lost between the components of a system in a quantum superposition. In terms of density matrices, the loss of interference effects corresponds to the diagonalization of the "environmentally traced over" density matrix. That is, the observed folding rate would be the average of the single-torsion transition rate (*N* =1) over different torsional degrees of freedom. We have used this decoherence model to calculate the protein folding rate. It is found that the parameter $\varepsilon$ calculated in this model is certainly lower than the value required by experiments. The ratio of the maximum permissible $\varepsilon$ calculated in decoherence model, $\varepsilon_{dec}^{\max}$, to that required by experiments for 16 proteins are listed in Table 1. Apart from one protein 1l2y(wt), $\varepsilon_{dec}^{\max}$ is lower than the value deduced from experiments by a factor 0.2-0.4 for most proteins. This indicates there exist some difficulties in explaining the temperature-dependence of folding rate by use of single-torsion transition. The multi-torsion coherence model seems better than the decoherence model.

    Moreover, the multi-torsion coherence model can explain not only the curious non-Arrhenius temperature-dependencies of the folding for a given protein, but also the specific statistical distribution of the folding rates for all measured two-state proteins. It was proved that in multi-torsion coherence model the analytical formula on folding rate Eq (74) is consistent with all existing experimental data of 65 two-state proteins and the correlation between theoretical prediction with experimental folding rate has attained 74% to 78% [17] [23]. Recently，Garbuzynskiy et al indicated that the measured protein folding rates fall within a narrow triangle (called Golden triangle)[24]. The Golden triangle can be interpreted by our quantum folding model.

    Why the multi-torsion coherent quantum theory can interpret the protein folding experiments successfully, without regard to the decoherence effect of environment? The decoherence effect is estimated by computing the decoherence time of the molecular system under thermal environment. The rigorous solution of decoherence time is difficult but some simple models were proposed. One such model introduced in [22] proved the decoherence time $\tau_D = \tau_R (\frac{\hbar}{\Delta x \sqrt{2mk_B T}})^2$ where

$\tau_R$ is the relaxation time and $\frac{\hbar}{\sqrt{2mk_BT}}$ the thermal de Broglie wavelength, $m$ the particle mass and $\Delta x$ the dimension of particle. It leads to $\tau_D = 10^2 \tau_R$ for an electron but about $10^{-2} \tau_R$ for an atom (carbon), much shorter than the electronic decoherence time. To study the decoherence effect on protein folding, we generalize the above model to the torsional angular motion of the macromolecule and obtain an estimate of torsional decoherence

$$\tau_D = \tau_R (\frac{\hbar}{\Delta\theta\sqrt{2Ik_BT}})^2 > 10^4 \tau_R (\frac{\hbar}{\sqrt{2Ik_BT}})^2 \cong \tau_R \quad (86)$$

where $\Delta\theta$ is the uncertainty of torsional angle and $I$ the inertia moment of atomic group. Notice that $\sqrt{2Ik_BT} = J_{therm}$ is the thermal angular momentum and $\Delta\theta$ satisfies $\Delta\theta \Delta J \approx \hbar$ ($\Delta J$ is the uncertainty of angular momentum). Eq (86) can be rewritten as $\tau_D \approx \tau_R(\frac{\Delta J}{J_{therm}})^2$. In the last inequality of Eq (86), $\Delta\theta \leq 0.5$ degree (about one tenth of the angular shift $\delta\theta$ in torsion potential) and $I = 10^{-37} \text{g} \cdot \text{cm}^2$ have been taken. By comparing the torsion decoherence with electronic and atomic decoherence time we find, if the relaxation rates are same in three cases then the decoherence effect on molecular torsion is in the mid of electron and atom. So, even if the quantum coherence for the macromolecule as a particle may have been destroyed the torsion-coherence of a protein can still exist. This gives a possible explanation on the success of the multi-torsion coherence model in studying protein folding.

To summarize, since protein molecules are the building blocks of living organism and they function properly everywhere in the life the proposed folding rate equations may have a broad application in molecular biological problems. The fast variables possibly include the electron coordinates, the stretching/bending of single bond, the atomic group or ligand binding to polypeptide chain, the hydrophobicity interaction between subunits, and the isomerization under external action etc. All these fast variables are slaved by slow variables — the torsional vibration and torsional transition of the protein. The existence of a set of coherent quantum oscillators of torsion-vibration type of low frequency and the quantum transition between torsion states — this forms a universal mechanism for a large class problems of the conformational change in molecular biology. We have studied the temperature dependencies of the protein folding and the overall statistical distribution of folding rates for all two-state proteins [17][18]. Other applications include the generalization of two-state protein to multi-state protein folding [25], the conformational transition rate of tubulin dimer in microtubules[23], the ligand binding in G protein-coupled receptors in membrane, the histone modification in nucleic acid through atomic group binding [12], and the protein photo-folding processes (the photon emission or absorption in protein folding and the inelastic scattering of photon on protein) [26]. The preliminary studies on these conformational changes from the point of quantum transition have been worked out and the proposed quantum folding theory will be tested further.

**Acknowledgement**   The author is indebted to Dr. Lu Jun for his detailed statistical analyses of protein experimental data.  He would like to thank Dr Zhao Judong, Dr Zhang Ying and Dr Zhang Lirong for numerous discussions in the writing of the paper.

**Figure legends**

**Figure 1  U(θ) – θ relation for the j-th torsion mode**. The frequency parameters for initial and final potential are $\omega_j$ and $\omega'_j$ respectively. The energy gap between initial and final states $\delta E_j$ and the angular shift $\delta\theta_j$ are labeled. For simplicity the subscript $j$ has been omitted here.

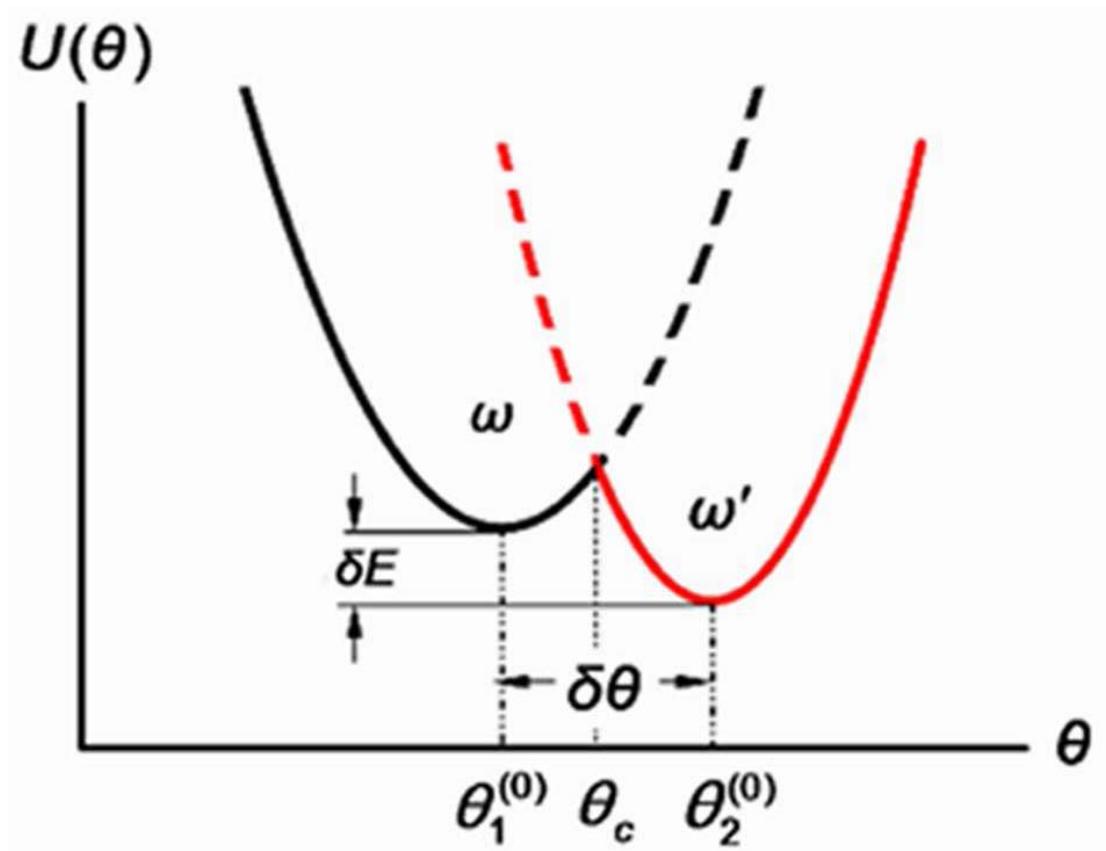

**Table 1. Temperature dependence and torsion parameters for 16 proteins**

| PDB code | S | R | $\bar{\omega}$ ($\times 10^{12}$ Hz) | N | $N_{th}$ | $\varepsilon_{dec}^{max}/\varepsilon$ |
|---|---|---|---|---|---|---|
| 1bdd | -24669 | 0.2441 | 1.4153 | 212 | 8 | 0.132 |
| 1div | -32076 | 0.2930 | 0.9649 | 229 | 4 | 0.264 |
| 1e0l | -16241 | 0.1780 | 1.5995 | 97 | 5 | 0.227 |
| 1enh | -33182 | 0.3345 | 0.8424 | 227 | 3 | 0.349 |
| 1iet | -70322 | 0.7322 | 0.5463 | 346 | 2 | 0.545 |
| 1l2y(p12w) | -14796 | 0.1602 | 1.4398 | 73 | 3 | 0.372 |
| 1l2y(wt) | -18957 | 0.1774 | 0.7023 | 73 | 1 | 1.562 |
| 1lmb(wt) | -83920 | 0.7613 | 0.7684 | 307 | 4 | 0.310 |
| 1lmb(g46a) | -30292 | 0.3313 | 0.6155 | 307 | 3 | 0.484 |
| 1lmb(sa37g) | -112897 | 1.0766 | 1.4154 | 307 | 11 | 0.091 |
| 1pin(wt) | -69675 | 0.6812 | 1.1677 | 129 | 4 | 0.320 |
| 1pin(s18g) | -77113 | 0.7565 | 1.0819 | 128 | 3 | 0.375 |
| 1pin(n26d) | -27063 | 0.2990 | 0.9520 | 129 | 3 | 0.481 |
| 1prb | -47886 | 0.3893 | 1.3737 | 179 | 7 | 0.166 |
| 2a3d | -18486 | 0.1812 | 1.0186 | 273 | 6 | 0.199 |
| 2pdd | -159403 | 1.5560 | 0.7321 | 152 | 2 | 0.690 |

The PDB code for each studied protein is given in column 1. S and R are best-fit slope parameters of the folding temperature dependence, $\bar{\omega}$ is the torsion frequency calculated from S, R and N (the average torsion inertial moment of atomic groups in polypeptide $\langle I_j \rangle = I_0 = 10^{-44}$ kgm$^2$ and the average angular shift in the torsion potential $\delta\theta = 0.1$ are assumed). N is the number of torsion modes of the polypeptide chain calculated from Eq (75). All above calculations in columns 2-5 can be found in [17]. $N_{th}$ is the threshold of N as the torsion potential satisfies the condition $\omega < \omega_{max} = 7.5 \times 10^{12}$ sec$^{-1}$. $\varepsilon_{dec}^{max}$ is the maximum permissible $\varepsilon$ calculated in decoherence model which is explicitly lower than the experimental value for all proteins in the table except 1l2y(wt). In calculating $\varepsilon_{dec}^{max}$, N=1 and $\omega = \omega_{max}$ are taken.